\documentclass[%
 reprint,
superscriptaddress,
 amsmath,amssymb,
  aip
]{revtex4-2}

\usepackage{graphicx}
\usepackage{dcolumn}
\usepackage{bm}
\usepackage{siunitx}
\usepackage{hyperref}
\usepackage{placeins}
\usepackage[T1]{fontenc}
\usepackage{etoolbox}
\usepackage{xcolor}
\usepackage{mathptmx}
\usepackage{nameref}
\usepackage{booktabs}
\usepackage{chemformula}
\usepackage{ulem}

\hypersetup{
        colorlinks   = true,
        citecolor    = blue,
        linkcolor    = blue}

\usepackage{graphicx}
\usepackage{dcolumn}
\usepackage{bm}
\usepackage{lipsum}
\usepackage{siunitx}
\usepackage{hyperref}
\usepackage{placeins}
\usepackage{etoolbox}
\usepackage{mathptmx}
\usepackage{xcolor}
\usepackage{nameref}
\usepackage{booktabs}

\newcommand{\rev}[1]{\textcolor{black}{#1}}

\DeclareSIUnit\bar{bar}

\DeclareSIUnit\bar{bar}
\newcommand{\MBT}{$\text{Mn}\text{Bi}_{2}\text{Te}_{4}$ }

\begin{document}


\title{Hall-effect in the \MBT crystal using silicon nitride nanomembrane via contacts}


\author{Mickey Martini}
 \affiliation{Leibniz Institute for Solid State and Materials Research Dresden (IFW Dresden), 01069 Dresden, Germany}
 \affiliation{Institute of Applied Physics, Technische Universität Dresden, 01062 Dresden, Germany}

 \author{Tommaso Confalone}
 \affiliation{Leibniz Institute for Solid State and Materials Research Dresden (IFW Dresden), 01069 Dresden, Germany}
 \affiliation{Institute of Applied Physics, Technische Universität Dresden, 01062 Dresden, Germany}
 
  \author{Yejin Lee}
 \affiliation{Leibniz Institute for Solid State and Materials Research Dresden (IFW Dresden), 01069 Dresden, Germany}
  
\author{Bastian Rubrecht}
 \affiliation{Leibniz Institute for Solid State and Materials Research Dresden (IFW Dresden), 01069 Dresden, Germany}

 \author{Giuseppe Serpico}
 \affiliation{Department of Physics, University of Naples Federico II, 80125 Naples, Italy}

 \author{Sanaz Shokri}
 \affiliation{Leibniz Institute for Solid State and Materials Research Dresden (IFW Dresden), 01069 Dresden, Germany}

\author{Christian~N.~Saggau}
 \affiliation{Leibniz Institute for Solid State and Materials Research Dresden (IFW Dresden), 01069 Dresden, Germany}
 \author{Domenico Montemurro}
 \affiliation{Department of Physics, University of Naples Federico II, 80125 Naples, Italy}

  \author{Valerii~M.~Vinokur}
 \affiliation{Terra Quantum AG, 9400 Rorschach, Switzerland}
\affiliation{Physics Department, CUNY, The City College of New York, 160 Convent Avenue, 10031 New York, USA}

 \author{Anna Isaeva}
 \affiliation{Van der Waals – Zeeman Institute, IoP, University of Amsterdam, 1098 XH Amsterdam, The Netherlands}
 

 \author{Kornelius Nielsch}
 \affiliation{Leibniz Institute for Solid State and Materials Research Dresden (IFW Dresden), 01069 Dresden, Germany}
 \affiliation{Institute of Materials Science, Technische Universität Dresden, 01062 Dresden, Germany}
 
   \author{Nicola Poccia}
     \email{n.poccia@ifw-dresden.de}
 \affiliation{Leibniz Institute for Solid State and Materials Research Dresden (IFW Dresden), 01069 Dresden, Germany}

\date{\today}

\begin{abstract}
\noindent
Utilizing an interplay between band topology and intrinsic magnetism, the two-dimensional van der Waals (vdW) system \MBT provides an ideal platform for realizing exotic quantum phenomena and offers great opportunities in the emerging field of antiferromagnetic spintronic technology. Yet, the fabrication of MnBi\textsubscript{2}Te\textsubscript{4}-based nanodevices is hindered by the high sensitivity of this material, which quickly degrades when exposed to air or to elevated temperatures. Here, we demonstrate an \rev{alternative} route of fabricating vdW-MnBi\textsubscript{2}Te\textsubscript{4}-based electronic devices using the cryogenic dry transfer of a printable circuit embedded in an inorganic silicon nitride membrane. The electrical connections between the thin crystal and the top surface of the membrane are established through via contacts. 
Our magnetotransport study reveals that this innovative via contact approach enables exploring the MnBi\textsubscript{2}Te\textsubscript{4}-like sensitive 2D materials and engineering synthetic heterostructures as well as complex circuits based on the two-dimensional vdW systems. \end{abstract}


\maketitle

\onecolumngrid

~~ 

\twocolumngrid

 \begin{figure*}[t!]
    \centering
    \includegraphics[width=1\textwidth]{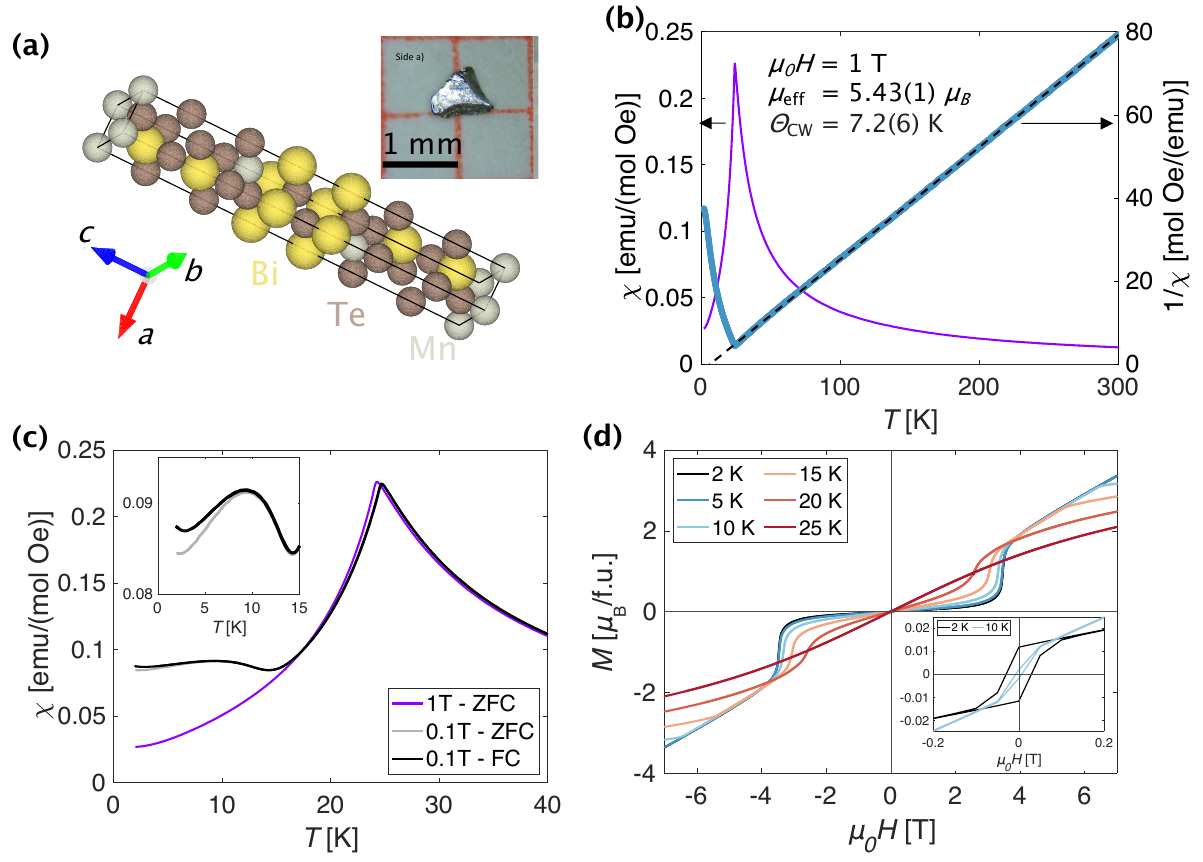}
    \caption{\textbf{(a)} Diagram of \MBT crystal structure and a synthesized single crystal. \textbf{(b)} Magnetic susceptibility (purple curve) and its reciprocal (blue curve) versus temperature under a magnetic field of $\SI{1}{\tesla}$. The parameters $\Theta_\mathrm{CW}$ and $\mu_\mathrm{eff}$ are, respectively, the Curie-Weiss temperature and effective moment used for the fitting dotted line. \textbf{(c)} Magnetic susceptibility versus temperature measured with a field of $\SI{1}{\tesla}$ and $\SI{0.1}{\tesla}$ after a zero field cooling, and with $\SI{0.1}{\tesla}$ after a field cooling of $\SI{0.1}{\tesla}$. Inset: Closeup of the two curves with $\SI{0.1}{\tesla}$ in the temperature range 0-$\SI{15}{\kelvin}$. \textbf{(d)}\,Magnetization as a function of a field sweep measured at different temperatures. Inset: Closeup of the hysteresis loop appearing at low fields for $\SI{2}{\kelvin}$ and $\SI{10}{\kelvin}$ temperatures.}
    \label{fig:SQUID}
\end{figure*}
Introducing magnetic order in a time-reversal-invariant topological insulator (TI) results in a formation of a non-trivial gap at the Dirac point, which brings in novel physics, in particular, the quantum anomalous Hall effect (QAHE)\,\cite{haldane1988model,hasan2010colloquium,yu2010quantized,qi2011topological,chang2013experimental,kou2014scale,checkelsky2014trajectory,bestwick2015precise,wang2015quantum,he2018topological, deng2020quantum}. The central phenomenon brought in by the QAHE is the dissipationless spin-polarized currents flowing along the edges of the material. This effect makes magnetic TIs a promising platform for topological magnetoelectronics having low dissipation losses\,\cite{he2022topological}. Recently, the van der Waals (vdW) \MBT has been predicted\,\cite{li2019intrinsic,zhang2019topological,otrokov2019prediction} and has been experimentally demonstrated\,\cite{gong2019experimental,otrokov2019prediction, lee2019spin,deng2020quantum,liu2020robust, ge2020high, bernevig2022progress} to be an antiferromagnetic TI. The \MBT compound consists of the Te-Bi-Te-Mn-Te-Bi-Te layers, where the Mn\textsuperscript{2+} magnetic moments are ferromagnetically aligned within each septuple layer (SL) and are   antiferromagnetically coupled when lying in neighboring SLs below Néel temperature $T_{\mathrm N}\simeq\SI{24}{\kelvin}$\,\cite{he2018topological}. The A-type AFM phase with the out-of-plane easy axis in \MBT determines its topological character. Depending on whether the \MBT flake has an even or odd number of SLs, the magnetizations of the top and bottom surfaces either sum up or fully compensate, resulting in an axion insulator or Chern insulator state, respectively\,\cite{li2019intrinsic, zhang2019topological,liu2020robust,ovchinnikov2021intertwined}. In the latter case, the QAHE is expected to arise below $T_{\mathrm N}$ when an appropriate gate voltage is applied to shift the Fermi energy close to the charge neutrality point \cite{deng2020quantum}. Nevertheless, the surface magnetic orders, and, hence, the topological properties of MnBi\textsubscript{2}Te\textsubscript{4}, can be easily broken by disorder and surface degradation, hindering the realization of the QAHE\,\cite{li2023progress}. A more robust way to open the surface gap and induce the chiral edge states is sandwiching \MBT between bidimensional FM insulators in order to decrease the interlayer AFM coupling and stabilize an FM order by proximity effects\,\cite{fu2020exchange}.  
Furthermore, because of the AFM structure and the low dimensionality, \MBT offers additional opportunities for the next generation of spintronic technologies. Magnetotransport studies have revealed that \MBT hosts the surface and bulk-spin flop transitions, resulting in the appearance of the 
magnetic phase that exhibits a chiral Hall effect and in the onset of the ferromagnetic/paramagnetic state emerging usually under applied strong magnetic fields\,\cite{sass2020robust,bac2022topological}. Very recently, \MBT was proposed as a novel material for engineering field-effect transistors possessing a strong rectifying effect and a spin filtering effect\,\cite{an2021nanodevices}. By devising twisted heterostructures through the stacking of the \MBT thin crystals, a Moiré magnetization texture emerges similar to what was observed in twisted CrI\textsubscript{3} bilayers\,\cite{song2021direct}, potentially leading to chiral channels networks\,\cite{xiao2020chiral}. 

To keep the disorder under control and realize the pristine interfaces in the proposed heterostructures, advanced device fabrication methods are needed. It is well known that the \MBT compound is not stable in the air since an oxide layer forms, passivating the surface of \MBT\,\cite{akhgar2022formation}. Moreover, \MBT decomposes into the non-stoichiometric Bi\textsubscript{2}Te\textsubscript{3} and MnTe at around $\SI{150}{\celsius}$, as revealed by X-ray diffraction experiments\,\cite{lee2013crystal,zeugner2019chemical}, and defects as well a shift of the chemical potential are induced by heating above \SI{120}{\celsius}\,\cite{breunig2022opportunities}. Therefore, the preservation of the physical properties of a MnBi\textsubscript{2}Te\textsubscript{4}-based crystal can be achieved through nanofabrication occurring in an entirely inert environment, with electrical contacts established in a dry and cold fashion. To minimize interfacial detrimental disorder associated with highly energetic metallization processes, several transfer techniques exploiting different materials have been proposed for the integration of vdW crystals in two-dimensional electronics\,
\cite{cui2015multi,chuang2016low,liu2016pushing,liu2018approaching,wang2019van,liu2019van,liu2021transferred,liu2022graphene,telford2018via,jung2019transferred}. Nevertheless, all these methods are based on non-scalable exfoliated materials and are typically limited to simple contact geometries\,\cite{purdie2018cleaning}. Only very recently, a wafer scale vdW integration process was developed to realize high-performance 2D transistors with high-quality contacts\,\cite{yang2023highly}. This approach relies on the direct transfer of electrical contacts embedded in poly(methyl methacrylate) (PMMA) layer onto the two-dimensional crystal, thereby separating the circuit fabrication stage from the realization of the device. Nonetheless, due to the high temperature required for melting the polymer after the contact printing, this methodology turns out to be unsuitable for sensitive material such as \MBT.


Here, we report on a via contact method that utilizes the cryogenic dry transfer of electrical circuits embedded within inorganic SiN\textsubscript{x} nanomembranes for the fabrication of sensitive vdW devices. We employ this experimental technique to realize and measure a Hall bar device based on a thin \MBT crystal. We show that this technology offers a unique opportunity for devising functional vdW heterostructure based on the TIs\,\cite{ovchinnikov2022topological} that are introduced into complex circuits. Our experimental approach can also be extended to other sensitive vdW materials, including high-temperature cuprate superconductors, which will allow for the realization of heterostructures and devices with the tunable properties\,\cite{martini2023twisted,lee2023encapsulating}, including topological superconductivity\,\cite{zhao2021emergent}.

 \begin{figure*}[t!]
    \centering
    \includegraphics[width=1\textwidth]{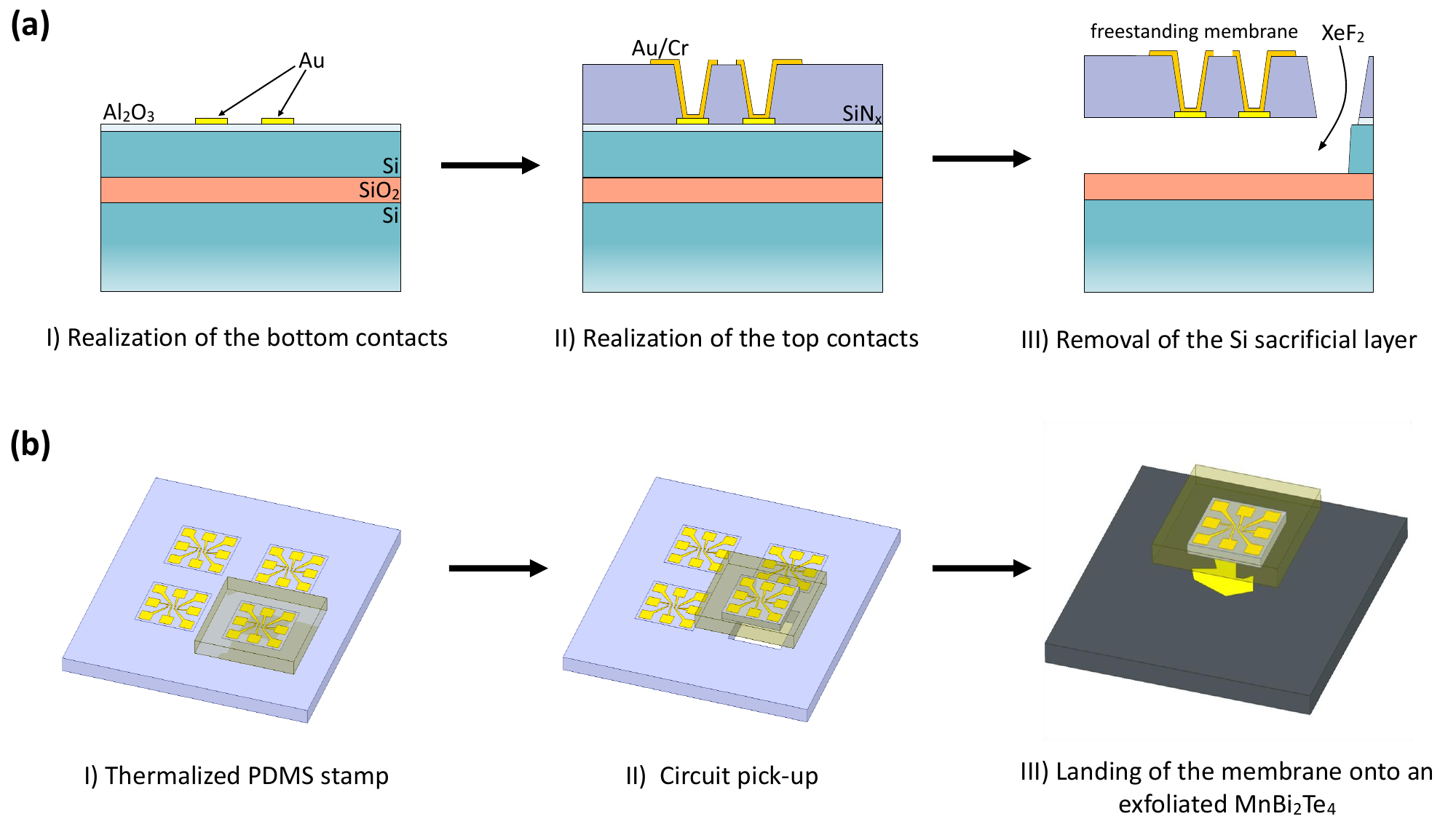}
    \caption{\textbf{(a)} Key steps for the nanomembrane fabrication. I) Coating of the substrate with Al\textsubscript{2}O\textsubscript{3} film and realization of the bottom contacts by sputtered Au. II) Deposition of the SiN\textsubscript{x} layer, etching the via electrodes, and creation of the electrical connection between the bottom and the top surface by sputtered Au/Cr. III) Definition of a physical pathway through which the XeF\textsubscript{2} gas can flow and selectively remove the sacrificial Si layer beneath each membrane, making them freestanding. \textbf{(b)} Printing of the electrical circuit onto the 2D crystal. I) Thermalization of the PDMS stamp in contact with the nanomembrane embedding the circuit at $-\SI{30}{\celsius}$. II) Picking up the nanomembrane by quickly detaching the stamp. III) Alignment and landing of the nanomembrane onto the fresh exfoliated MnBi\textsubscript{2}Te\textsubscript{4} crystal.}
    \label{fig:fabrication}
\end{figure*}

The high quality single crystals of the \MBT [Fig.~\hyperref[fig:SQUID]{1(a)}] are grown by slow crystallization from a melt of stoichiometric Bi\textsubscript{2}Te\textsubscript{3} and $\alpha$-MnTe. The homogenized precursor is placed inside an evacuated quartz tube. After sealing the ampule it is heated up to $\SI{950}{\celsius}$ in a tube furnace. The single crystals form during slow cooling in the narrow Ostwald-Miers region ($\SI{3}{\celsius}$ below $\SI{600}{\celsius}$) and are tempered at the subsolidification temperature $\SI{590}{\celsius}$ for several days. Finally, the temperature is lowered to room temperature. The platelet-shaped crystals of different sizes depending on the cooling rate and annealing time are obtained within and atop of the highly crystalline ingot\,\cite{zeugner2019chemical}.

 \begin{figure*}[t!]
    \centering
    \includegraphics[width=.9\textwidth]{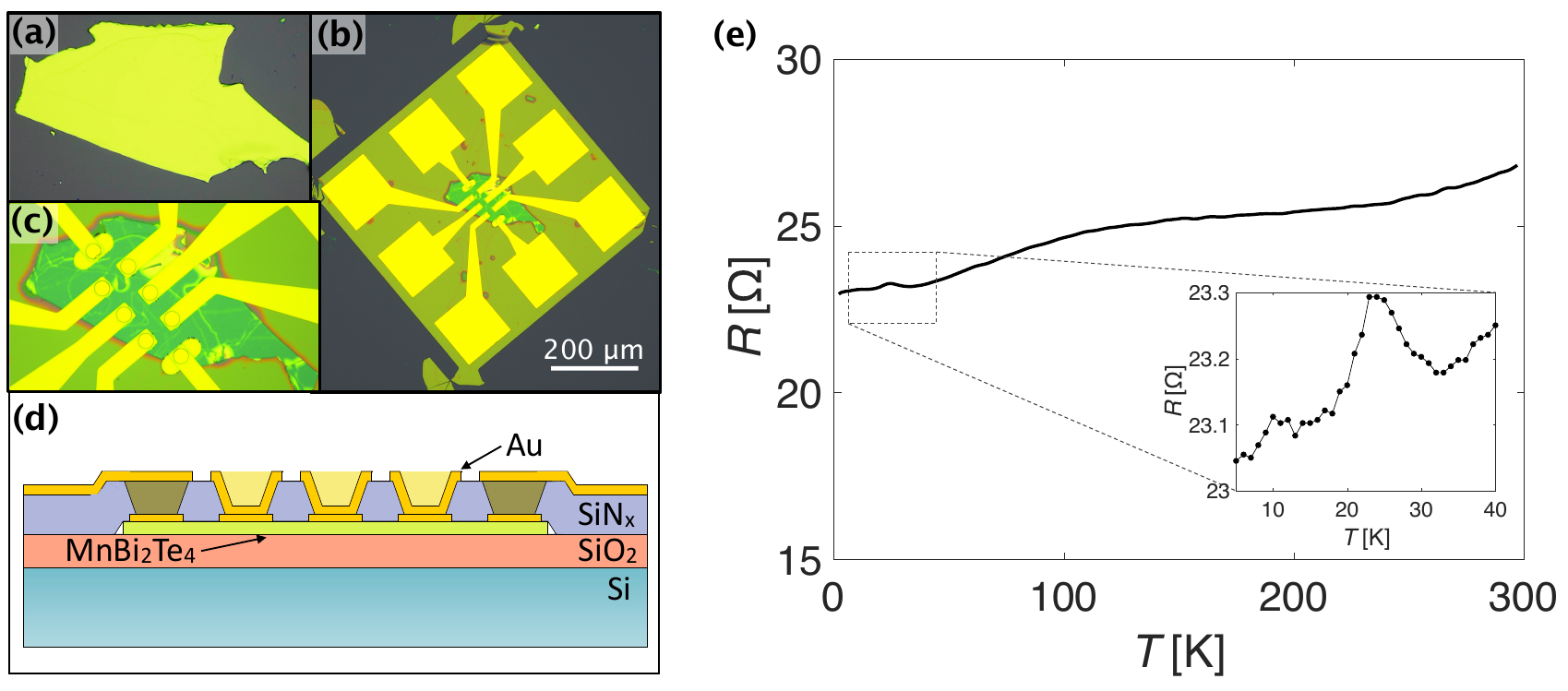}
    \caption{\textbf{(a)} Optical image of a \MBT flake investigated in this study. \textbf{(b)} Microscopy image of the full device with contacts printed circuit right before wire bonding. \textbf{(c)} Closeup image of the contact region. \textbf{(d)} Sketch of a cut through the central part of the membrane illustrating the via and bottom contacts. \textbf{(e)} Electrical resistance versus temperature. Inset: The resistance as a function of temperature in the interval 5-$\SI{40}{\kelvin}$.}
    \label{fig:device}
\end{figure*} 

The magnetometric measurements are performed in a superconducting quantum interference device under a magnetic field applied along the $c$-axis of the crystal. The magnetic susceptibility $\chi$ as a function of temperature [Fig.~\hyperref[fig:SQUID]{1(b)}] in a $\SI{1}{\tesla}$-field reveals an AFM transition at $T_{\mathrm N}= \SI{24.25}{\kelvin}$. The inverse susceptibility data are well fitted by the Curie-Weiss (CW) law $1/\chi = (T-\Theta_\mathrm{CW})/C$, where $C$ is the Curie constant connected to the effective magnetic moment of Mn-ion $\mu_\mathrm{eff}$, and $\Theta_\mathrm{CW}$ is the CW temperature, related to the sum of all magnetic interaction in the system. Due to Mn/Bi intersite defects appearing when the Mn atoms from the Mn layer intermix with the Bi sites in the Bi\textsubscript{2}Te\textsubscript{3} layers, the obtained $\mu_\mathrm{eff}=5.43(1)~\mu_{\mathrm B}$/f.u. lies slightly below the expected value for a pure Mn$^{2+}$ $\mu_\mathrm{eff}^{Mn2+}=5.9~\mu_{\mathrm B}$/Mn\,\cite{lai2021defect,liu2021site,garnica2022native}. The observed peak in the magnetic susceptibility indicates an AFM order, while the extrapolated positive $\Theta_\mathrm{CW} = +\SI{7.2\pm0.6}{\kelvin}$ suggests the predominant ferromagnetic correlations. This contradiction is well known in the layered vdW compounds like MnBi\textsubscript{2}Te\textsubscript{4}. Here the ferromagnetic intralayer coupling is much stronger than the antiferromagnetic interlayer coupling so that the spin fluctuations in the paramagnetic regime are dominated by the FM correlations. The magnetic susceptibility measured at a lower field ($H = \SI{.1}{\tesla}$) shows a peak at a slightly higher temperature ($T_{\mathrm N}=\SI{24.75}{\kelvin}$), and discloses a second transition at the critical temperature $T_{\mathrm c} \simeq \SI{10}{\kelvin}$ [Fig.~\hyperref[fig:SQUID]{1(c)}], which coincides with the ferromagnetic ordering temperature of Mn\textsubscript{x}Bi\textsubscript{2-x}Te\textsubscript{3}\,\cite{hor2010development}. This additional phase is also visible in the field-dependent magnetization $M(H)$ curves at low temperatures. As shown in the inset of Fig.~\hyperref[fig:SQUID]{1(d)}, a hysteresis loop appears at low fields and vanishes for $T>\SI{10}{\kelvin}$. However, from the remanent moment of $\simeq 0.01~\mu_{\mathrm B}$/f.u. it can be seen that the fraction of this secondary phase is negligibly small ($< 1\%$), in agreement with previous studies\,\cite{yan2019crystal}. Figure\,\hyperref[fig:SQUID]{1(d)} also reveals a spin-flop transition occurring at around $\mu_0H=\SI{3.5}{\tesla}$ and $T=\SI{2}{\kelvin}$, when the field is applied along the crystallographic $c$-axis, that leads the system to a canted AFM state. At $\SI{20}{\kelvin}$, the transition is much more broadened and appears around $\mu_0H=\SI{2.5}{\tesla}$, while it completely disappears above $T_{\mathrm N}$.

To study the magneto-transport in thin \MBT, we employ a via contacts technique that enables the realization of a Hall bar device entirely in an argon environment without exposure to heat or chemicals. This methodology relies on the contact printing of a SiN\textsubscript{x} membrane, hosting a microcircuit, onto the thin vdW crystal. The latter is contacted by bottom contacts which are connected through the via contacts with the bonding pads at the top surface of the membrane. Therefore, our experimental technique completely separates the formation of the device's electrical connections from the actual fabrication of these electrical contact lines using the standard clean room processes.

An array containing about a hundred membranes is structured on a $\SI{1}{\centi\meter}\times\SI{1}{\centi\meter}$ Si/SiO\textsubscript{2}/Si substrate, where the top sacrificial silicon layer allows for the creation of freestanding transferable circuits. The fabrication of the membranes starts by coating the substrate with a $\SI{5}{\nano\meter}$-thick Al\textsubscript{2}O\textsubscript{3} film by atomic layer deposition. The bottom electrical contacts are defined by standard optical or e-beam lithography and lift-off process with $\SI{80}{\nano\meter}$ of sputtered Au, as illustrated in Fig.~\hyperref[fig:fabrication]{2(a)\,I)}. Next, the chip is covered by a $\SI{500}{\nano\meter}$-thick SiN\textsubscript{x} layer, deposited by the plasma-enhanced chemical vapor deposition. The via electrodes connecting the bottom and the top surfaces of the membranes are formed by lithography and reactive ion etching processes. These and the top pads are formed by sputtering $\SI{5}{\nano\meter}$ of Cr and $\SI{80}{\nano\meter}$ of Au, as depicted in Fig.~\hyperref[fig:fabrication]{2(a)\,II)}. Finally, a physical pathway through which the XeF\textsubscript{2} gas can flow and selectively remove the sacrificial Si layer beneath each membrane is obtained by means of etching processes [Fig.~\hyperref[fig:fabrication]{2(a)\,III)}]. After the action of the XeF\textsubscript{2} gas, the membranes become freestanding. \rev{The details of the nanomembrane fabrication procedure can be found in Ref.~\cite{Saggau2023mem}.}

The preparation of the device in an inert atmosphere starts with placing the chip hosting the transferable circuits on a liquid nitrogen-cooled stage kept at $-\SI{30}{\celsius}$. A PDMS stamp is brought into contact with a nanomembrane utilizing a micromanipulator and let it thermalize to enhance its stickiness, as sketched in Fig.~\hyperref[fig:fabrication]{2(b)\,I}. By quickly detaching the stamp from the chip, we pick up the target membrane hosting the circuit, as shown in Fig.~\hyperref[fig:fabrication]{2(b)\,II}. Using mechanical exfoliation, we isolate a \rev{$\SI{120}{\nano\meter}$-thick} \MBT flake on a Si$\mathrm{O}_{2}$/Si, see Fig.~\hyperref[fig:fabrication]{2(b)}\,III, previously treated with oxygen plasma to enhance the vdW forces between the crystal and the substrate. Finally, the nanomembrane is transferred and freely released from the stamp on top of the flake at room temperature, as shown in Fig.~\hyperref[fig:fabrication]{2(b)\,IV}. \rev{The flexibility of the membranes scales inversely with their thickness; therefore, membranes with a thickness below $\SI{500}{\nano\meter}$ cannot be easily transferred to the substrate utilizing a conventional PDMS stamp.} The \MBT crystal and the resulting device before the wire bonding are illustrated in Figs.~\hyperref[fig:device]{3(a-c)}. Figure.~\hyperref[fig:device]{3(d)} shows the schematic drawing of the cross-section of the device.
With this dry transfer technique, we achieve a high-quality electrical contact (areal contact resistance  $<\SI{50}{\kilo\ohm\micro\meter^2}$). \rev{The stability of the electronic properties of sensitive devices realized with nanomembranes via contacts is discussed in Ref.~\cite{Saggau2023mem}.} Figure \hyperref[fig:device]{3(e)} shows the temperature dependence of the longitudinal resistance measured in a four-point geometry. Two main features are apparent in this data set. First, the electrical resistance shows a metallic behavior with the change in slope in the range $150-\SI{200}{\kelvin}$, as already observed in\,Ref.~\cite{rani2019crystal}. Next, a pronounced peak arises at $T_{\mathrm N}=\SI{25}{\kelvin}$ due to the enhanced spin scattering across the AFM transition. On the other hand, electrical resistance does not show any signature of additional phases observed in the susceptibility data around\,$\SI{10}{\kelvin}$. Unlike magnetization measurements, the magnetotransport properties are much less sensitive to magnetic impurities.  

 \begin{figure}[t!]
    \centering
    \includegraphics[width=.45\textwidth]{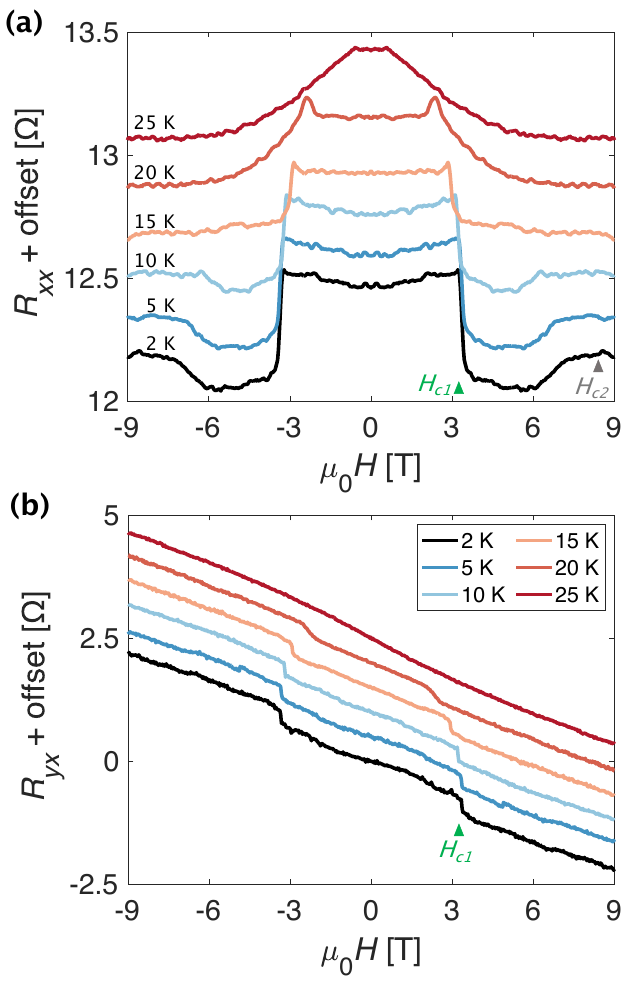}
    \caption{\textbf{(a)} The MR under out-of-plane magnetic field at various temperatures. \textbf{(b)} Magnetic-field dependence of Hall resistance measured at the same temperatures.}
    \label{fig:transport}
\end{figure}
 To quantify the magnetoresistance (MR) and Hall effect, we inject an AC current $i=\SI{50}{\micro\ampere}$ along the $x$-axis and measure the longitudinal, $V_\mathrm{xx}$, and transversal, $V_\mathrm{yx}$, voltages using a lock-in amplifier technique while sweeping an out-of-plane magnetic field in both directions. The longitudinal and transversal resistances correspond, respectively, to the diagonal and off-diagonal components of the resistance tensor. Respectively, their variations are typically even and odd functions of the magnetic field. We take advantage of this property to symmetrize the longitudinal voltage and antisymmetrize the transversal voltage with respect to the magnetic field. This way, we can decouple the longitudinal MR and Hall effects, which are otherwise weakly mixed up due to the geometry of the electric field which is not exactly parallel to the $x$-axis, possibly, due to an irregular shape of the \MBT flake. 
 
 The MR curves measured in an out-of-plane magnetic field at different temperatures are shown in Fig.~\hyperref[fig:transport]{4(a)}. Below $T_{\mathrm N}$, the MR sharply drops at the critical field $H_\mathrm{c1}$ where the spin-flop transition occurs, and the established spin order leads to an AFM state. For $T\leq\SI{10}{\kelvin}$, upon further increase of the magnetic field, the MR curve forms a kink at $H_\mathrm{c2}$ before starting to decrease again since spins get polarized to an FM state with the suppressed spin fluctuations\,\cite{liu2020robust}. At $T=\SI{10}{\kelvin}$ this occurs around $H=\SI{6.5}{\tesla}$, exactly where the magnetization in Fig.~\hyperref[fig:SQUID]{1(c)} starts to saturate, whereas at lower temperatures it occurs at higher fields, $H>\SI{7}{\tesla}$; therefore a comparison with the SQUID data is not available. On the other hand, at higher temperatures, the spin canting is smeared out over a broad field range, resulting in a less sharp and monotonous drop of the MR. The curves displaying the Hall resistance $R_\mathrm{yx} = V_\mathrm{yx}/i$  versus magnetic field are measured at the same temperatures and displayed in Fig.~\hyperref[fig:transport]{4(b)}. The negative slope of $R_\mathrm{yx}(H)$ demonstrates that the charge transport is of the electron type. From the linear region of $R_\mathrm{yx}$, we estimate the carrier concentration $n_0=\SI{2.8e20}{cm^{-3}}$ and the carrier mobility $\mu=\SI{180}{cm^2\per\volt\per\second}$ at $T=\SI{2}{\kelvin}$; $n_0=\SI{2.4e20}{cm^{-3}}$ and $\mu=\SI{200}{cm^2\per\volt\per\second}$ at  $T=\SI{20}{\kelvin}$. The carrier mobility is much smaller than that of pure Bi\textsubscript{2}Te\textsubscript{3} ($\sim\SI{1000}{cm^2\per\volt\per\second}$)\,\cite{holgado2020shubnikov,stephen2020weak}, in agreement with previous studies\,\cite{otrokov2019prediction, liu2020robust}. This can arise because of the non-negligible amount of Mn/Bi antisites and cation vacancies\,\cite{zeugner2019chemical}. Nevertheless, the $R_\mathrm{yx}$ traces do not display any hysteretic behavior, suggesting that the intermixing of Mn and Bi does not dramatically affect the magnetotransport in the \MBT flake. 


In summary, we demonstrate an \rev{alternative} route to integrating sensitive van der Waals crystals like \MBT into nanodevices. This integration is achieved by the cryogenic dry transfer of microcircuits embedded into an inorganic silicon nitride nanomembrane, where the via contacts connect the bottom electrical contacts at the 2D crystal with the top surface of the membrane. By employing this method, we realize a Hall bar of \MBT in an entirely inert environment without exposing the material to chemicals or elevated temperatures. The fabricated device displays high-quality electrical contacts having resistance below $\SI{50}{\kilo\ohm\per\micro\meter^2}$. Our magnetotransport study reveals an AFM phase transition around $\SI{25}{\kelvin}$, supporting the susceptibility measurements performed in the \MBT bulk crystal. Systematic measurements of the magnetoresistance and Hall effect at different temperatures confirm the existence of the spin flop transitions at the intermediate magnetic fields, followed by a canted magnetic phase, in agreement with the field dependence of the magnetization of the bulk crystal. All these findings suggest that this via contact technique can be exploited for engineering more complex nanodevices based on sensitive materials. 

\medskip

\noindent
\textbf{Data availability}
The data that support the findings of this study are available from the corresponding author upon reasonable request.\\

\medskip

\noindent\\~~
\textbf{Acknowledgements} The experiments were partially supported by the Deutsche Forschungsgemeinschaft (DFG 452128813, DFG 512734967, DFG 492704387, DFG 460444718). The work of V.M.V was supported by Terra Quantum AG and partly by the NSF award 2105048. The authors are grateful to Heiko Reith and Ronny Engelhard for providing access to cleanroom infrastructure and to Nicolas Perez Rodriguez for providing access to cryogenic measurement facilities. 

\medskip
\noindent\\~~
\textbf{Author contributions.} N.P. designed and supervised the experiment; C.N.S. and S.S. fabricated the nanomembranes; M.M, Y.L, S.S. developed the cryogenic transfer methodology; M.M., Y.L., B.R. performed the experiments and analyzed the data with the contribution of T.C.; the \MBT crystals have been provided by A.I.; N.P. and V.M.V. discussed the results. The manuscript has been written by M.M., T.C., B.R., V.M.V., K.N., and N.P. All authors discussed the manuscript.

\medskip
\noindent
\textbf{Conflict of Interest} All authors declare no conflict of interest. 

\bibliographystyle{}


\begin{thebibliography}{}
\bibitem{deng2020quantum}Deng, Y., Yu, Y., Shi, M., Guo, Z., Xu, Z., Wang, J., Chen, X. \& Zhang, Y. Quantum anomalous Hall effect in intrinsic magnetic topological insulator \MBT. {\em Science}. \textbf{367}, 895-900 (2020)
\bibitem{chang2013experimental}Chang, C., Zhang, J., Feng, X., Shen, J., Zhang, Z., Guo, M., Li, K., Ou, Y., Wei, P., Wang, L. \& Others Experimental observation of the quantum anomalous Hall effect in a magnetic topological insulator. {\em Science}. \textbf{340}, 167-170 (2013)
\bibitem{wang2015quantum}Wang, J., Lian, B. \& Zhang, S. Quantum anomalous Hall effect in magnetic topological insulators. {\em Phys. Scr}. \textbf{2015}, 014003 (2015)
\bibitem{kou2014scale}Kou, X., Guo, S., Fan, Y., Pan, L., Lang, M., Jiang, Y., Shao, Q., Nie, T., Murata, K., Tang, J. \& Others Scale-invariant quantum anomalous Hall effect in magnetic topological insulators beyond the two-dimensional limit. {\em Phys. Rev. Lett}. \textbf{113}, 137201 (2014)
\bibitem{he2018topological}He, K., Wang, Y. \& Xue, Q. Topological materials: quantum anomalous Hall system. {\em Annu. Rev. Condens. Matter Phys}. \textbf{9} pp. 329-344 (2018)
\bibitem{hasan2010colloquium}Hasan, M. \& Kane, C. Colloquium: topological insulators. {\em Rev. Mod. Phys}. \textbf{82}, 3045 (2010)
\bibitem{qi2011topological}Qi, X. \& Zhang, S. Topological insulators and superconductors. {\em Rev. Mod. Phys}. \textbf{83}, 1057 (2011)
\bibitem{haldane1988model}Haldane, F. Model for a quantum Hall effect without Landau levels: Condensed-matter realization of the" parity anomaly". {\em Phys. Rev. Lett}. \textbf{61}, 2015 (1988)
\bibitem{yu2010quantized}Yu, R., Zhang, W., Zhang, H., Zhang, S., Dai, X. \& Fang, Z. Quantized anomalous Hall effect in magnetic topological insulators. {\em Science}. \textbf{329}, 61-64 (2010)
\bibitem{checkelsky2014trajectory}Checkelsky, J., Yoshimi, R., Tsukazaki, A., Takahashi, K., Kozuka, Y., Falson, J., Kawasaki, M. \& Tokura, Y. Trajectory of the anomalous Hall effect towards the quantized state in a ferromagnetic topological insulator. {\em Nat. Phys}. \textbf{10}, 731-736 (2014)
\bibitem{bestwick2015precise}Bestwick, A., Fox, E., Kou, X., Pan, L., Wang, K. \& Goldhaber-Gordon, D. Precise quantization of the anomalous Hall effect near zero magnetic field. {\em Phys. Rev. Lett}. \textbf{114}, 187201 (2015)
\bibitem{he2022topological}He, Q., Hughes, T., Armitage, N., Tokura, Y. \& Wang, K. Topological spintronics and magnetoelectronics. {\em Nat. Mater.}. \textbf{21}, 15-23 (2022)
\bibitem{li2019intrinsic}Li, J., Li, Y., Du, S., Wang, Z., Gu, B., Zhang, S., He, K., Duan, W. \& Xu, Y. Intrinsic magnetic topological insulators in van der Waals layered \MBT-family materials. {\em Sci. Adv}. \textbf{5}, eaaw5685 (2019)
\bibitem{zhang2019topological}Zhang, D., Shi, M., Zhu, T., Xing, D., Zhang, H. \& Wang, J. Topological axion states in the magnetic insulator \MBT with the quantized magnetoelectric effect. {\em Phys. Rev. Lett}. \textbf{122}, 206401 (2019)
\bibitem{otrokov2019prediction}Otrokov, M., Klimovskikh, I., Bentmann, H., Estyunin, D., Zeugner, A., Aliev, Z., Gaß, S., Wolter, A., Koroleva, A., Shikin, A. \& Others Prediction and observation of an antiferromagnetic topological insulator. {\em Nature}. \textbf{576}, 416-422 (2019)
\bibitem{gong2019experimental}Gong, Y., Guo, J., Li, J., Zhu, K., Liao, M., Liu, X., Zhang, Q., Gu, L., Tang, L., Feng, X. \& Others Experimental realization of an intrinsic magnetic topological insulator. {\em Chin. Phys. Lett}. \textbf{36}, 076801 (2019)
\bibitem{lee2019spin}Lee, S., Zhu, Y., Wang, Y., Miao, L., Pillsbury, T., Yi, H., Kempinger, S., Hu, J., Heikes, C., Quarterman, P. \& Others Spin scattering and noncollinear spin structure-induced intrinsic anomalous Hall effect in antiferromagnetic topological insulator \MBT. {\em Phys. Rev. Res}. \textbf{1}, 012011 (2019)
\bibitem{liu2020robust}Liu, C., Wang, Y., Li, H., Wu, Y., Li, Y., Li, J., He, K., Xu, Y., Zhang, J. \& Wang, Y. Robust axion insulator and Chern insulator phases in a two-dimensional antiferromagnetic topological insulator. {\em Nat. Mater}. \textbf{19}, 522-527 (2020)
\bibitem{ge2020high}Ge, J., Liu, Y., Li, J., Li, H., Luo, T., Wu, Y., Xu, Y. \& Wang, J. High-Chern-number and high-temperature quantum Hall effect without Landau levels. {\em Natl. Sci. Rev.}. \textbf{7}, 1280-1287 (2020)
\bibitem{bernevig2022progress}Bernevig, B., Felser, C. \& Beidenkopf, H. Progress and prospects in magnetic topological materials. {\em Nature}. \textbf{603}, 41-51 (2022)
\bibitem{ovchinnikov2021intertwined}Ovchinnikov, D., Huang, X., Lin, Z., Fei, Z., Cai, J., Song, T., He, M., Jiang, Q., Wang, C., Li, H. \& Others Intertwined topological and magnetic orders in atomically thin Chern insulator \MBT. {\em Nano Lett}. \textbf{21}, 2544-2550 (2021)
\bibitem{li2023progress}Li, S., Liu, T., Liu, C., Wang, Y., Lu, H. \& Xie, X. Progress on antiferromagnetic topological insulator \MBT. {\em Natl. Sci. Rev.}. \textbf{186}, 227-236 (2017)
\bibitem{fu2020exchange}Fu, H., Liu, C. \& Yan, B. Exchange bias and quantum anomalous Hall effect in the \MBT/CrI\textsubscript{3} heterostructure. {\em Sci. Adv}. \textbf{6}, eaaz0948 (2020)
\bibitem{sass2020robust}Sass, P., Kim, J., Vanderbilt, D., Yan, J. \& Wu, W. Robust a-type order and spin-flop transition on the surface of the antiferromagnetic topological insulator \MBT. {\em Phys. Rev. Lett}. \textbf{125}, 037201 (2020)
\bibitem{bac2022topological}Bac, S., Koller, K., Lux, F., Wang, J., Riney, L., Borisiak, K., Powers, W., Zhukovskyi, M., Orlova, T., Dobrowolska, M. \& Others Topological response of the anomalous Hall effect in \MBT due to magnetic canting. {\em npj Quantum Mater}. \textbf{7}, 1-7 (2022)
\bibitem{an2021nanodevices}An, Y., Wang, K., Gong, S., Hou, Y., Ma, C., Zhu, M., Zhao, C., Wang, T., Ma, S., Wang, H. \& Others Nanodevices engineering and spin transport properties of \MBT monolayer. {\em NPJ Comput. Mater}. \textbf{7}, 45 (2021)
\bibitem{song2021direct}Song, T., Sun, Q., Anderson, E., Wang, C., Qian, J., Taniguchi, T., Watanabe, K., McGuire, M., Stöhr, R., Xiao, D. \& Others Direct visualization of magnetic domains and moiré magnetism in twisted 2D magnets. {\em Science}. \textbf{374}, 1140-1144 (2021)
\bibitem{xiao2020chiral}Xiao, C., Tang, J., Zhao, P., Tong, Q. \& Yao, W. Chiral channel network from magnetization textures in two-dimensional \MBT. {\em Phys. Rev. B}. \textbf{102}, 125409 (2020)
\bibitem{akhgar2022formation}Akhgar, G., Li, Q., Di Bernardo, I., Trang, C., Liu, C., Zavabeti, A., Karel, J., Tadich, A., Fuhrer, M. \& Edmonds, M. Formation of a Stable Surface Oxide in \MBT Thin Films. {\em ACS Appl. Mater. Interfaces}. \textbf{14}, 6102-6108 (2022)
\bibitem{zeugner2019chemical}Zeugner, A., Nietschke, F., Wolter, A., Gaß, S., Vidal, R., Peixoto, T., Pohl, D., Damm, C., Lubk, A., Hentrich, R. \& Others Chemical aspects of the candidate antiferromagnetic topological insulator \MBT. {\em Chem. Mater}. \textbf{31}, 2795-2806 (2019)
\bibitem{lee2013crystal}Lee, D., Kim, T., Park, C., Chung, C., Lim, Y., Seo, W. \& Park, H. Crystal structure, properties and nanostructuring of a new layered chalcogenide semiconductor, Bi\textsubscript{2}MnTe\textsubscript{4}. {\em CrystEngComm}. \textbf{15}, 5532-5538 (2013)
\bibitem{breunig2022opportunities}Breunig, O. \& Ando, Y. Opportunities in topological insulator devices. {\em Nat. Rev. Phys}. \textbf{4}, 184-193 (2022)
\bibitem{cui2015multi}Cui, X., Lee, G., Kim, Y., Arefe, G., Huang, P., Lee, C., Chenet, D., Zhang, X., Wang, L., Ye, F. \& Others Multi-terminal transport measurements of MoS2 using a van der Waals heterostructure device platform. {\em Nat.~Nanotech.}. \textbf{10}, 534-540 (2015)
\bibitem{chuang2016low}Chuang, H., Chamlagain, B., Koehler, M., Perera, M., Yan, J., Mandrus, D., Tomanek, D. \& Zhou, Z. Low-resistance 2D/2D ohmic contacts: a universal approach to high-performance WSe\textsubscript{2}, MoS\textsubscript{2}, and MoSe\textsubscript{2} transistors. {\em Nano Lett.}. \textbf{16}, 1896-1902 (2016)
\bibitem{liu2016pushing}Liu, Y., Guo, J., Wu, Y., Zhu, E., Weiss, N., He, Q., Wu, H., Cheng, H., Xu, Y., Shakir, I. \& Others Pushing the performance limit of sub-100 nm molybdenum disulfide transistors. {\em Nano Lett.}. \textbf{16}, 6337-6342 (2016)
\bibitem{liu2018approaching}Liu, Y., Guo, J., Zhu, E., Liao, L., Lee, S., Ding, M., Shakir, I., Gambin, V., Huang, Y. \& Duan, X. Approaching the Schottky–Mott limit in van der Waals metal–semiconductor junctions. {\em Nature}. \textbf{557}, 696-700 (2018)
\bibitem{wang2019van}Wang, Y., Kim, J., Wu, R., Martinez, J., Song, X., Yang, J., Zhao, F., Mkhoyan, A., Jeong, H. \& Chhowalla, M. Van der Waals contacts between three-dimensional metals and two-dimensional semiconductors. {\em Nature}. \textbf{568}, 70-74 (2019)
\bibitem{liu2019van}Liu, Y., Huang, Y. \& Duan, X. Van der Waals integration before and beyond two-dimensional materials. {\em Nature}. \textbf{567}, 323-333 (2019)
\bibitem{liu2021transferred}Liu, L., Kong, L., Li, Q., He, C., Ren, L., Tao, Q., Yang, X., Lin, J., Zhao, B., Li, Z., Chen, Y., Li, W., Song, W., Lu, Z., Li, G., Li, S., Duan, X., Pan, A., Liao, L. \& Liu, Y. Transferred van der Waals metal electrodes for sub-1-nm $\text{Mo}\text{S}_{2}$ vertical transistors. {\em Nat. Electron}. \textbf{4} pp. 342-347 (2021)
\bibitem{liu2022graphene}Liu, G., Tian, Z., Yang, Z., Xue, Z., Zhang, M., Hu, X., Wang, Y., Yang, Y., Chu, P., Mei, Y., Liao, L., Hu, W. \& Di, Z. Graphene-assisted metal transfer printing for wafer-scale integration of metal electrodes and two-dimensional materials. {\em Nat. Electron}. \textbf{5} pp. 275-280 (2022)
\bibitem{telford2018via}Telford, E., Benyamini, A., Rhodes, D., Wang, D., Jung, Y., Zangiabadi, A., Watanabe, K., Taniguchi, T., Jia, S., Barmak, K., Pasupathy, A., Dean, C. \& Hone, J. Via Method for Lithography Free Contact and Preservation of 2D Materials. {\em Nano Lett}. \textbf{18} pp. 1416-1420 (2018)
\bibitem{jung2019transferred}Jung, Y., Choi, M., Nipane, A., Borah, A., Kim, B., Zangiabadi, A., Taniguchi, T., Watanabe, K., Yoo, W., Hone, J. \& Teherani, J. Transferred via contacts as a platform for ideal two-dimensional transistors. {\em Nat. Electron}. \textbf{2} pp. 187-194 (2019)
\bibitem{purdie2018cleaning}Purdie, D., Pugno, N., Taniguchi, T., Watanabe, K., Ferrari, A. \& Lombardo, A. Cleaning interfaces in layered materials heterostructures. {\em Nat.~Comm.}. \textbf{9}, 5387 (2018)
\bibitem{yang2023highly}Yang, X., Li, J., Song, R., Zhao, B., Tang, J., Kong, L., Huang, H., Zhang, Z., Liao, L., Liu, Y. \& Others Highly reproducible van der Waals integration of two-dimensional electronics on the wafer scale. {\em Nature Nanotechnology}. \textbf{18}, 471-478 (2023)
\bibitem{ovchinnikov2022topological}Ovchinnikov, D., Cai, J., Lin, Z., Fei, Z., Liu, Z., Cui, Y., Cobden, D., Chu, J., Chang, C., Xiao, D. \& Others Topological current divider in a Chern insulator junction. {\em Nat. Commun}. \textbf{13}, 5967 (2022)
\bibitem{martini2023twisted}Martini, M., Lee, Y., Confalone, T., Shokri, S., Saggau, C., Wolf, D., Gu, G., Watanabe, K., Taniguchi, T., Montemurro, D. \& Others Twisted cuprate van der Waals heterostructures with controlled Josephson coupling. {\em Mater Today}. , \textbf{67}, 106-112 (2023)
\bibitem{lee2023encapsulating}Lee, Y., Martini, M., Confalone, T., Shokri, S., Saggau, C., Wolf, D., Gu, G., Watanabe, K., Taniguchi, T., Montemurro, D. \& Others Encapsulating High-Temperature Superconducting Twisted van der Waals Heterostructures Blocks Detrimental Effects of Disorder. {\em Adv. Mater}. \textbf{35}, 2209135 (2023)
\bibitem{zhao2021emergent}Zhao, S., Poccia, N., Cui, X., Volkov, P., Yoo, H., Engelke, R., Ronen, Y., Zhong, R., Gu, G., Plugge, S. \& Others Emergent interfacial superconductivity between twisted cuprate superconductors. {\em ArXiv.org ArXiv:2108.13455}. (2021)
\bibitem{garnica2022native}Garnica, M., Otrokov, M., Aguilar, P., Klimovskikh, I., Estyunin, D., Aliev, Z., Amiraslanov, I., Abdullayev, N., Zverev, V., Babanly, M. \& Others Native point defects and their implications for the Dirac point gap at \MBT (0001). {\em npj Quantum Mater}. \textbf{7}, 1-9 (2022)
\bibitem{liu2021site}Liu, Y., Wang, L., Zheng, Q., Huang, Z., Wang, X., Chi, M., Wu, Y., Chakoumakos, B., McGuire, M., Sales, B. \& Others Site mixing for engineering magnetic topological insulators. {\em Phys. Rev. X}. \textbf{11}, 021033 (2021)
\bibitem{lai2021defect}Lai, Y., Ke, L., Yan, J., McDonald, R. \& McQueeney, R. Defect-driven ferrimagnetism and hidden magnetization in MnBiTe. {Phys. Rev. B}. \textbf{103}, 184429 (2021)
\bibitem{hor2010development}Hor, Y., Roushan, P., Beidenkopf, H., Seo, J., Qu, D., Checkelsky, J., Wray, L., Hsieh, D., Xia, Y., Xu, S. \& Others Development of ferromagnetism in the doped topological insulator Bi\textsubscript{2-x}Mn\textsubscript{x}Te\textsubscript{3}. {\em Phys. Rev. B}. \textbf{81}, 195203 (2010)
\bibitem{yan2019crystal}Yan, J., Zhang, Q., Heitmann, T., Huang, Z., Chen, K., Cheng, J., Wu, W., Vaknin, D., Sales, B. \& McQueeney, R. Crystal growth and magnetic structure of \MBT. {\em Phys. Rev. Mater}. \textbf{3}, 064202 (2019)
\bibitem{Saggau2023mem}Saggau C.~N., Shokri, S., Martini, M., Confalone, T., Lee, Y., Wolf, D., Gu, G., Brosco, V., \& Others, 2D High-Temperature Superconductor Integration in Contact Printed Circuit Boards
 {\em ACS Appl Mater Interfaces}. \textbf{15}, 51558 (2023)
\bibitem{rani2019crystal}Rani, P., Saxena, A., Sultana, R., Nagpal, V., Islam, S., Patnaik, S. \& Awana, V. Crystal growth and basic transport and magnetic properties of \MBT. {\em J. Supercond. Nov. Magn}. \textbf{32}, 3705-3709 (2019)
\bibitem{holgado2020shubnikov}Holgado, D., Bolaños, K., Castro, S., Monteiro, H., Pena, F., Okazaki, A., Fornari, C., Rappl, P., Abramof, E., Soares, D. \& Others, Shubnikov–de Haas oscillations and Rashba splitting in Bi\textsubscript{2}Te\textsubscript{3} epitaxial film. {\em Appl. Phys. Lett}. \textbf{117}, 102108 (2020)
\bibitem{stephen2020weak}Stephen, G., Vail, O., Lu, J., Beck, W., Taylor, P., Friedman, A. \& Others Weak antilocalization and anisotropic magnetoresistance as a probe of surface states in topological Bi\textsubscript{2}Te\textsubscript{x}Se\textsubscript{3-x} thin films. {\em Sci. Rep}. \textbf{10}, 1-7 (2020)
















\end{thebibliography}

\end{document}